\documentclass[10pt,letterpaper,twocolumn]{article} 

\usepackage{ol2}
\usepackage[draft]{hyperref}
\usepackage{amsmath}

\begin{document}

\twocolumn[ 

\title{Coupling of single quantum emitters to plasmons propagating on mechanically etched wires}

\author{Shailesh Kumar,$^{*}$ Alexander Huck, Ying-Wei Lu, and Ulrik L. Andersen}

\address{Department of Physics, Technical University of Denmark, Building 309, 2800 Lyngby, Denmark}

\email{$^{*}$shailesh@fysik.dtu.dk}

\begin{abstract}
We demonstrate the coupling of a single nitrogen vacancy center in a nanodiamond  to propagating plasmonic modes of mechanically etched silver nanowires. The mechanical etch is performed on single crystalline silver nanoplates by the tip of an atomic force microscope cantilever to produce wires with pre-designed lengths. We show that single plasmon propagation can be obtained in these wires, thus making these structures a platform for quantum information processing.   
\end{abstract}

 ] 

Surface plasmon modes on metallic nanostructures can confine optical fields to dimensions much smaller than the vacuum wavelength of the associated light field. This offers the possibility of miniaturizing photonic circuits~\cite{2010Bozhevolnyi, 2010Schuller}. The high confinement also enables strong coherent interaction of individual quantum emitters with single mode fields~\cite{2006Chang, 2010Bozhevolnyi, 2010Schuller, 2007Akimov, 2009Kolesov,2011Huck, 2013NLShailesh, 2013APLShailesh}. However, as a result of the strong field confinement and thus field enhancement, the requirements on the fabrication of the nanostructures become very critical. The standard approach for obtaining the desired metallic nanostructures has been electron-beam lithography followed by thermal evaporation of metal~\cite{Refereedemand}. This method, however, results in polycrystalline metallic nanostructures which leads to severe scattering from the grain boundaries, resulting in high propagation losses~\cite{PRLAgCavity}. The polycrystalline nature of metallic structure also gives rise to a strong fluorescence which makes it  less suitable for coupling to single quantum emitters such as nitrogen vacancy (NV) centers in diamond ~\cite{ArxivTopDown}.

In order to avoid scattering losses, single-crystalline nanowires of silver and gold with exquisite propagation properties have been fabricated using chemical synthesis~\cite{Auwirefabrication, Agwirefabrication}. However, the drawback of this fabrication method is that  structures cannot be predesigned~\cite{Auplatefabrication, MicrosizedAgcrystals}. Recently, a different method that combines bottom-up and top-down approaches has been applied to fabricate complex, single-crystalline nanostructures of gold and silver~\cite{2010Hecht, Goldflakemachine,2012Shailesh}. Single-crystalline silver nanoplates (SNPs) were produced using chemical synthesis and, subsequently, the desired structure was formed by the use of focused ion beam (FIB) milling~\cite{2012Shailesh}. Using this method, we have fabricated very smooth structures and shown that they support propagating plasmons~\cite{2012Shailesh}. However, this method requires the substrate to be conducting in order to avoid charging effects. The use of a conducting substrate leads to higher propagation losses for the plasmonic modes guided in the silver wire~\cite{2012JRKrenn}. In addition, Ga-ions used for FIB milling introduces defects in the structures.

In this letter, we demonstrate coupling of a single NV-center to a silver nanowire that are fabricated by mechanical etching of chemically synthesized single-crystalline SNPs. The etch is performed by scratching the SNPs with the tip of an atomic force microscope (AFM) cantilever. Propagation of plasmons is demonstrated in these nanowires. In addition, we couple a single quantum emitter to the plasmon modes of the wire which is witnessed through the measurements of the second order correlation function. 

SNPs were synthesized using the chemical process described in ref.~\cite{2012Shailesh}. The SNPs have heights in the range of 50-100 nm and an average area of around $100~\mu m^2$ and were spun onto a fused silica substrate. To fabricate and characterize the nanowires, we use a standard home built fluorescence confocal microscope in combination with an AFM. In the confocal microscope, green light at 532 nm is focused onto the structures and the fluorescence light is collected by a high numerical aperture (NA = 1.4) lens and directed through filters (transmits the wavelength range between 647 nm and 785 nm) to an avalanche photodiode (APD). The chemical process used for the synthesis of SNPs also produces a small amount of silver particles as a byproduct.  Silver particles are known to fluoresce when excited with a laser in the visible spectral region~\cite{Agfluorescence}. AFM topography images and fluorescence images are taken for SNPs before starting the fabrication process of nanowires. We fabricate nanowires in a region of an SNP where fluorescence is negligible, i.e. where the silver particles are not observed in the AFM image~\cite{2012Shailesh}. 

In materials with low scratch resistance, it is relatively easy to make grooves or peel away parts of the material using a tip with large hardness (or large scratch resistance). Depending on the force applied to the tip as well as the hardness of the soft material, multiple movements might be required to make a penetrating cut~\cite{PRLscratching}. We apply this principle to prepare silver nanowires by etching away parts of the single crystalline SNPs using an AFM tip with a large hardness. Using a similar technique, single crystalline wires have been fabricated from gold nanoplates, where diamond ultramicrotome is used to cut the plates embedded in epoxy~\cite{2008Nanoskiving}.

To carve out silver nanowires from SNPs, we either use diamond coated silicon cantilevers or bare silicon cantilevers. Diamond is known to have the highest hardness of 10 on the Mohs' ordinal scale of 0-10~\cite{Mohsscale}.  Silicon has a hardness of 7, and silver in the range of 2.4-4 in comparison. To make the nanowires, we press the substrate with the AFM cantilever in contact mode operation with a contact force of approximately $1~\mu N$. The area of SNP that we want to etch away is then scanned with a cantilever speed in the range of $10-15~\mu m/s$. Fewer scans are needed to cut the SNPs when diamond coated cantilevers are used. However, thinner wires are obtained when a silicon cantilever is used. This also removes most of the silver particles that are produced during the process. Some additional cleaning is needed to remove the remaining silver particles, which is done by scanning the area in contact mode operation of the AFM once again, but with a contact force nearly 10 times lower. 

\begin{figure}[htbp]
\begin{center}
\includegraphics[scale=1]{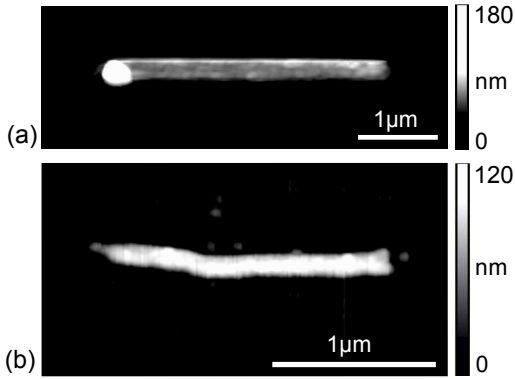}
\caption{\textbf{(a)} and  \textbf{(b)} AFM images of two different wires fabricated by mechanical etching of SNPs with a diamond coated cantilever and a silicon cantilever, respectively.  \label{fig:AFM_wires}}
\end{center}
\end{figure}

Figure~\ref{fig:AFM_wires} shows AFM images of two mechanically etched nanowires. Wires of different dimensions can be made; the height and the maximum length of the nanowires are however ultimately limited by the dimensions of the SNPs. The width of the nanowire is mainly limited by the adhesive force between the fabricated silver nanowire and the substrate. For the silica substrate that was used in our experiment, the minimum width of the wires was about 100~nm. The used SNPs have an area of around $100\mu m^2$, and the nanowires are produced by a mechanical etch of  the area surrounding the nanowires. Depending on whether a diamond coated or a silicon cantilever is used, it takes around 5 or 30 minutes to make a single silver nanowire, respectively. For making wires with a width larger than 300 nm, a diamond cantilever is used, and it takes either one or two scans to make the nanowires. The yield in this case is around 50 percent. Silicon cantilevers are used for making wires with widths less than 300 nm.  10-15 scans are needed to make these wires, and the yield is around 20 percent. In some cases, particles produced in the scratching process remained on the silver nanowires, as can be seen for example in Fig.~\ref{fig:AFM_wires}(a). 

\begin{figure}[htbp]
\begin{center}
\includegraphics[scale=1]{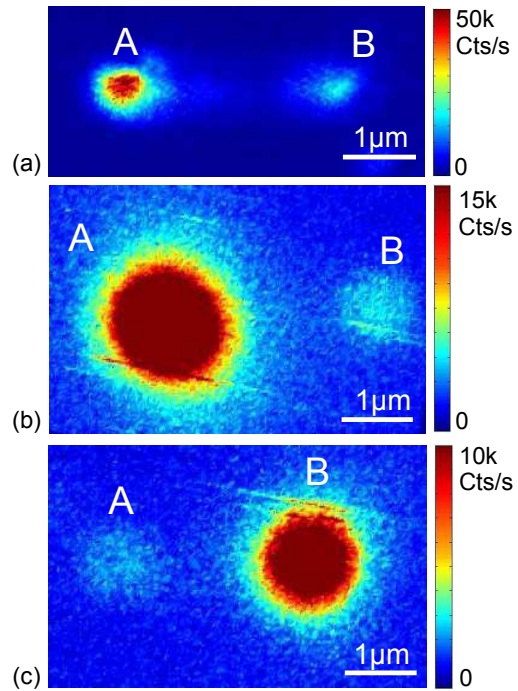}
\caption{\textbf{(a)} Fluorescence image of the silver nanowire whose AFM image is shown in Fig.~\ref{fig:AFM_wires}(a). \textbf{(b)} Galvanometric image when spot A in (a) is excited.  \textbf{(c)} Galvanometric image when spot B in (a) is excited. \label{fig:Propagation_AFMwires}}
\end{center}
\end{figure}

As a first test of the fabricated silver wires we investigate whether they support propagating plasmon modes. A fluorescence image is shown in Fig.~\ref{fig:Propagation_AFMwires}(a) for the wire whose AFM image is presented in Fig.~\ref{fig:AFM_wires}(a). The fluorescence image is obtained by a raster scan of the area and detection of fluorescence in the wavelength range between 647~nm and 785~nm. The fluorescence from the nanowire is not uniform. The ends of the wire fluoresce more due to the small silver particles that remain after the fabrication process. End \lq A\rq ~ fluoresces more than end  \lq B\rq ~as a result of the large silver particle attached to end \lq A\rq ~ as can be seen in the corresponding AFM image. The fluorescence from the middle region of the wire is low. We used fluorescence from the silver particles to excite the plasmonic modes propagating along the nanowire. Figure~\ref{fig:Propagation_AFMwires}(b) shows a fluorescence image taken by continuously exciting end \lq A\rq  ~of the wire (shown in Fig.~\ref{fig:Propagation_AFMwires}(a)) and scanning the plane containing the wire with a galvanometric mirror. In the galvanometric image, fluorescence from the distal end of the nanowire is observed. This means that the fluorescence from spot \lq A\rq ~in Fig.~\ref{fig:Propagation_AFMwires}(b) got coupled to propagating modes of the nanowire, it propagated to the distal end, and some of the plasmon polaritons got scattered into the far field. This appears as spot \lq B\rq ~in Fig.~\ref{fig:Propagation_AFMwires}(b). Similarly, a galvanometric scan image of fluorescence was taken while continuously exciting end \lq B\rq ~of the wire in Fig.~\ref{fig:Propagation_AFMwires}(a), in which case fluorescence from end \lq A\rq  ~was observed. This verifies the propagation of plasmon polaritons in AFM fabricated silver nanowires. 

\begin{figure}[htbp]
\begin{center}
\includegraphics[scale=1]{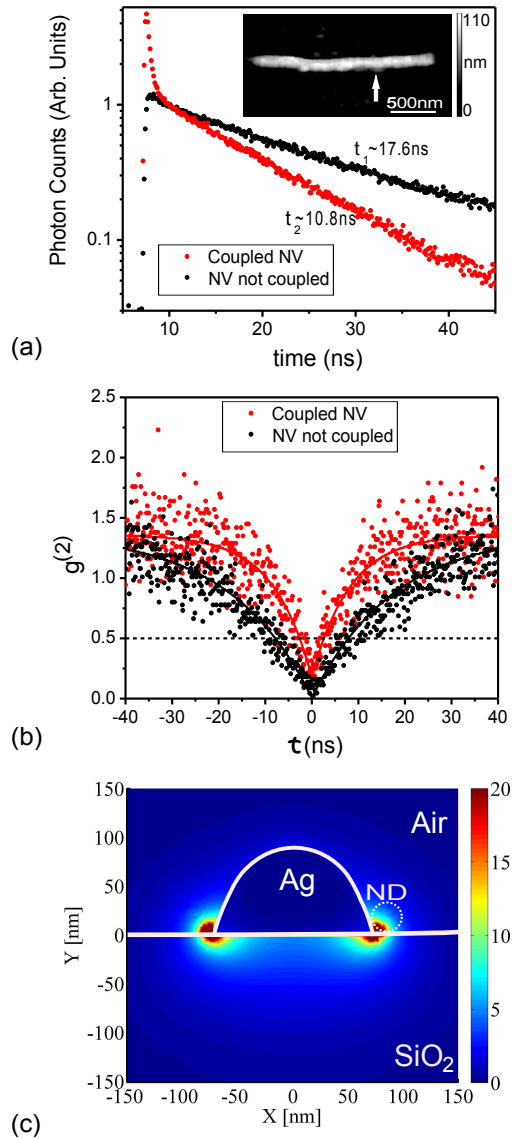}
\caption{ \textbf{(a)} Lifetime measurement data for an NV-center before and after coupling to a silver nanowire. The inset shows an AFM image of the coupled system where the white arrow indicates the position of the nanodiamond. \textbf{(b)} Autocorrelation measurement data for the NV-center before and after coupling to the silver nanowire. Fitted curves according to a three level model are also shown. \textbf{(c)} $\Gamma_{pl,max}/\Gamma_{0} $ plotted for the mode supported by the nanowire shown in the inset of (a). The most likely position of the nanodiamond (ND) is indicated by the dashed circle. \label{fig:NV_coupledwire}}
\end{center}
\end{figure}

We now use a mechanically etched nanowire for coupling a single NV-center in a nanodiamond to its plasmonic modes~\cite{2009Kolesov, 2011Huck}. To do this experiment, solutions containing nanodiamonds (size $< 50~nm$) and that containing SNPs were successively spin coated on a fused silica substrate. A silver nanowire was then fabricated, whose AFM image is presented in Fig. 1(b). An isolated nanodiamond, with height around 30~nm, containing a single NV-center was identified by performing measurements of its lifetime and autocorrelation function, which are presented in Fig.~\ref{fig:NV_coupledwire}. The lifetime of the NV-center is measured by exciting it with a pulsed laser (wavelength: 532nm, pulse width: 4.6ps, pulse rate: 5.05MHz) and measuring emission times with respect to the excitation times. The autocorrelation measurement is done using a standard Hanbury-Brown and Twiss set-up. The measured autocorrelation data is fitted with a curve according to a three level model for an NV-center~\cite{2000Kurtsiefer}.

After locating a nanocrystal diamond containing a single NV-center, we move the crystal across the sample so that it is placed within a few nanometers from the surface of the nanowire. An AFM image of the coupled system is shown in the inset of Fig.~3(a). The length of the nanowire is $1.8~\mu m$. The  nanowire height is 90~nm and the width of the nanowire is estimated to be around 150~nm. Figure~3(a) shows lifetimes of the NV center in the nanodiamond when it was far from the nanowire and when the nanodiamond was positioned close to the wire. The observed fast decay in the lifetime measurement after coupling of the NV-center to the nanowire is due to fluorescence from the silver particles created along the wire, whereas the slow decay component corresponds to the decay of the NV-center. From a comparison of the lifetimes, we estimate a rate enhancement of 1.63. We measured the NV-center's auto-correlations before and after it was coupled to the nanowire, and the results are shown in Fig. 3(b). It is clear that  $g^{(2)}(0)<0.5$ both for the coupled and uncoupled NV-center, which confirms single photon emission from the NV-center. The fluorescence from the silver wire is low which allowed us to measure the single photon character of the emission from an NV-center. Fluorescence from the wires can be reduced further with additional chemical cleaning.

We have calculated the coupling rate of a dipole emitter (emitting at vacuum wavelength of 700 nm) to a silver nanowire with a cross-section corresponding to the one shown in Fig. 3(a). We used a refractive index of 1.46 for the fused silica substrate and for silver we used the refractive index given in ref.  \cite{Palik}. First, we calculated the mode fields using a commercial software (COMSOL), and second, following ref.~\cite{PRBYuntian}, we deduced the coupling rate between an emitter and the resulting mode fields. Fig. 3(c) shows a plot of the coupling rate to the silver nanowire relative to the rate in vacuum maximized over all the dipole orientations ($\Gamma_{pl,max}/\Gamma_{0} $), as a function of the position in the plane of the silver wire cross-section. It can be seen from Fig. 3(c) that the coupling of the emitter to the silver nanowire crucially depends on the position of the NV-center inside the nanodiamond. It also depends on the orientation of the dipole with respect to the mode fields. To obtain higher rate enhancements, thinner silver wires or other silver structures such as grooves and gap structures could be considered. Making grooves and gap structures should, in principle, be possible with the technique described in this letter. Although, the placement of a nanodiamond crystal inside the groove or gap structures is a challenging task, it can be done with the aid of the pick-and-place technique~\cite{2011AFMpickup, 2011SEMpickup}.

In conclusion, we have fabricated single crystalline silver nanowires by combining bottom-up and top-down techniques that support travelling plasmon modes, and we have demonstrated coupling of these modes to a single NV center in diamond. Chemically synthesized SNPs were mechanically etched by an AFM cantilever to construct pre-designed nanowires, albeit with some limitations to the dimensions. Smaller transversal dimensions can be realized by increasing the adhesion power between the SNP and the substrate by using a polymer layer on the substrate, for example. It is also possible to make structures that are different from the linear ones made in this work by gaining higher control over the tip movement. Moreover, as an outlook it will be important to make a systematic study of the plasmonic propagation length for different dimensions of the wire and compare these results to other types of top-down fabricated nanowires.

The authors acknowledge financial support by the Villum Kann Rasmussen foundation and the Carlsberg foundation.

\end{document}